# A device to assess the upper limb isometric muscle strength

**Paulo José Oliveira Cortez**,
São Paulo State University (UNESP) / Department of Mechanics, Guaratinguetá – SP, Brazil,
paulocortez@universitas.edu.br

**José Elias Tomazini**,
São Paulo State University (UNESP) / Department of Mechanics, Guaratinguetá – SP, Brazil,
tomazini@feg.unesp.br

**Jorge Henrique O. Sales**,
Instituto de Física Teórica – UNESP, Rua Dr. Bento Teobaldo Ferraz 271, Bloco II, Barra Funda, São Paulo – SP - CEP: 01140-070,
jorgehenrique@unifei.edu.br

**Alfredo T. Suzuki**,
Instituto de Física Teórica – UNESP, Rua Dr. Bento Teobaldo Ferraz 271, Bloco II, Barra Funda, São Paulo – SP - CEP: 01140-070,
suzuki@ift.unesp.br

**Abstract**

In this work we project and build an equipment for upper limb muscle strength assessment and we did some assays to measure it from volunteers.

**Introduction**

The precise assessment of the performance of the human musculoskeletal system has been the objective of scientists and physical and rehabilitation medicine professionals for many decades (Perrin, 1993).

With the technological developments it turns out to be possible to quantify the performance of human beings. So, any assessment of technical sport, performance, functional capacity, among others, must be preceded by measurement, description and analysis (Winter, 1979).

According DVIR (2002) there is a lack of information about the functional connections of the upper limb and the factors involved in the production of force from this body region. In face of the necessity to obtain quantitative data in Biomechanics, one can perceive the need for projects and the development of equipments to measure the forces that interact with the locomotion system.

The aim for this project was to project and construct a device to measure in an efficacious manner the upper limb isometric muscle strength.

**Methods**

The definition of the model of the Station for Measuring Strength (SMS) was based on the position desired for the isometric strength tests. The placement allows the assessment of efforts (horizontal and vertical) with the subject in sitting position, and the SMS is composed of a common commercial chair and a Strength Measuring Device (SMD).

*Strength Measuring Device (SMD)*

The SMD is formed by an adjustable support structure for the forearm, accessory for placement of wrist joint and transducers. Chosen, as sensor to measure efforts, strain gauges duly pasted in a steel pipe and connected to an electrical circuit forming a complete Wheatstone bridge.

*Construction of the Strength Measuring Device*

The Strength Measuring Device (SMD) is formed by a steel tube instrumented with eight (8) strain gauges, trademark Kyowa, KFG-3-120-c1-11 model, and two (2) at the top and two (2) at the bottom of the tube to measure the isometric strength made in the vertical direction (flexion of the elbow and wrist joint). To measure the efforts made towards horizontal (internal and external rotation of the shoulder joint) were pasted two (2) gauges on each side portion of the steel tube. The strain gauges were glued onto each side of the lateral portion of the steel tube. The strain gauges were glued at 200 mm distance from the point of application of force by the subject, respecting the calculations performed earlier.

The Strength Measuring Device (SMD) has been dimensioned to support 500 N as a maximum load with $(\Delta E/V)_{máx} = 0,002$. The steel tube dimensions are as follows: external diameter $d_2 = 21,34$ mm, internal diameter $d_1 = 16,11$ mm and thickness $t = 2,74$ mm.

One of the tube extremities was enchased in a steel block and in the other extremity a connector of ½ inch elbow type with threads was glued. This connector made it possible to the SMD to measure efforts made in different directions through the instruments to support the wrist/hand complex that will be described later on. To house the sensor elements a 50 x 50 mm square metalon bar was used. Figure 1 illustrates the elements in our SMD.

Two accessories (handle) were built to furnish support to the wrist and hand complex of the individuals that would be submitted to strength tests. The form of these handles respect the anatomic conformation of hands and the direction of the applied effort.

A supporting device with anatomic contours for the forearm was placed onto the structure that houses the steel tube. Such device was part of the forearm support of a commercial chair. In its posterior part a pipeline was fixed to permit horizontal dislocation and a screw to fix the support. This can be used to evaluate the muscular force both for the left side upper limb as well as the right upper limb.

To allow horizontal dislocation of the SMD, a system of corner with equal 2 (two) inches tabs was fixed laterally on the sustentation base for the Station for Measuring Strength (SMS). Vertical positioning adjustments were possible thanks to a telescopic tubular system. The fixed tube at the lateral corner (fixed tube) presents an external diameter d2 = 44 mm and internal diameter $d_1$ = 38,5 mm whereas the base tube of the SMD (movable tube) has an external diameter $d_2$ = 38 mm.

*Device Calibration*

The electrical signals from the Wheatstone bridge circuit formed by the strain gauges were transmitted to a signal conditioning system, Spider 8 model (HBM, Darmstadt, Germany) and processed by software - Catman (version 3.1, release 3, 1997 - 2000). The calibration was done so through static loads applied in the pre-defined handle of SMD, while the direction of effort (vertical and horizontal). The results values (ΔE / V) were inserted in the Catman so that test of calibration (were converted automatically into Force (N). At the free end of the steel tube of the SMD, were applied the following masses: 1.98 [kg], 9.59 [kg], 17.22 [kg], 24.85 [kg], 32.45 [kg] 40.07 [kg], 47.68 [kg] and 55.31 [kg] which were measured on a balance accurate to load, corresponding respectively to the forces of 19.42 [N], 94.08 [N] , 168.93 [N], 243.78 [N], 318, 38 [N], 393.09 [N], 467.74 [N] and 542.59 [N].

**Results**

Figure 1 shows the SMD elements:

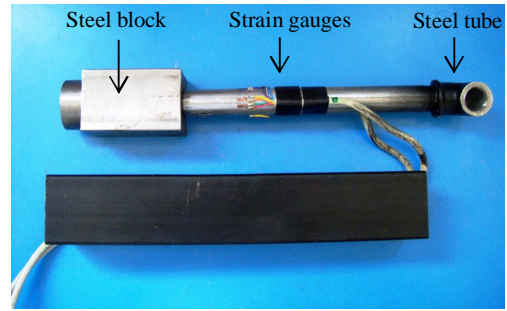

Figure 1 - Strength Measuring Device elements.

Figure 2 below illustrates the system of acquisition in its final stage

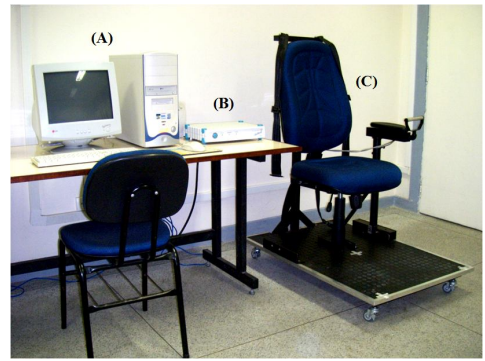

Figure 2 - (A) PC, (B) System for acquisition and conditioning signal SPIDER 8 model and (C) Strength Measuring Device (SMD).

Figure 3 illustrates the result of calibration for vertical efforts and the R-squared value. We observe an optimum linearity between the electric signal and the applied loads.

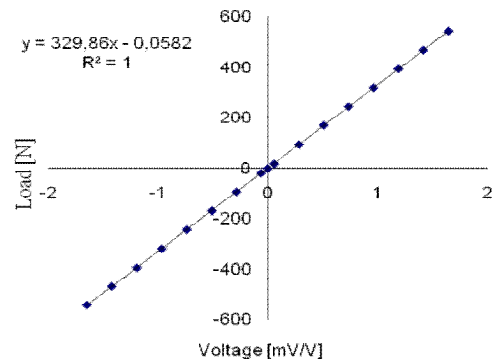

Figure 3 – Device Calibration data in vertical direction.

Figure 4 illustrates the result of calibration for horizontal efforts, the calibration equation and the R-squared value. Again we observe an excellent linearity between the electric signal and the applied loads. Negative values indicate change in the direction of force.

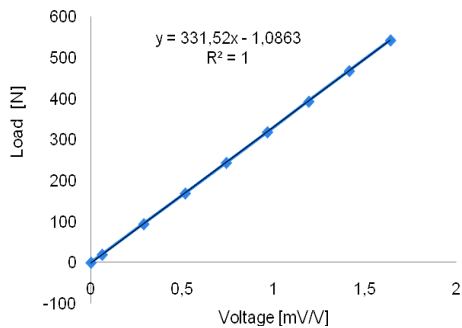

Figure 4 – Device Calibration data in the horizontal direction.

## Discussion

Analysis of muscle strength in an instrumental form is extensively made by researchers from commercially evaluated equipments such as computerized isokinetic dynamometers (Wilkin and Haddock, 2006; Salles, Filho, 2002; Hughes et al, 1999; Kasprisin, Grabiner, 2000), manual dynamometers (Bohannon, 1997; Pavan et al., 2006, Bohannon, 2004) and devices produced in specialized laboratories (Memberg *et al*., 2001; Ericson et al., 2002; Morse et al., 2006 ; Madeleine et al., 2006; De Groot et al. 2004; Garner and Shim, 2008).

Obtaining quantitative data in Biomechanics became a necessity within the academic-scientific scope and why not, clinic. From this, it is clear that there is a need for projects and developments of instruments to measure forces that interact with the locomotion system, as well as the low cost equipment availability for biomechanics analysis.

## Conclusion

The project, the construction, and the device calibration were successfully finalized steps, satisfying the researcher's expectations and the SMS is now in the adaptability, efficiency and experimental essays test phase.

Among the diverse options for future studies are the analysis of the force behavior in relation to time, muscle-skeleton fatigue, dominance and non-dominance relation, correlation between force and electromyography signal, obtaining of normative values for different populations, clinic evolution parameters in orthopedic and neurological disorders, among others.

## Acknowledgments

To FAPESP, FAPEMIG, CST (Companhia Siderúrgica de Tubarão), FUNDUNESP and to Instituto de Física Teórica-UNESP for the support to this project.